\IfFileExists{newtxtext.sty}{\documentclass{ametsocV6.1}}{\documentclass{analogtransportams}}
\usepackage[T1]{fontenc}
\usepackage[utf8]{inputenc}
\usepackage[letterpaper,margin=1in]{geometry}
\usepackage{booktabs,tabularx,array}
\usepackage{anyfontsize}
\usepackage{tikz}
\usetikzlibrary{arrows.meta,positioning,calc}
\usepackage{hyperref}
\hypersetup{hidelinks}
\usepackage{placeins}
\usepackage{docmute}
\usepackage{needspace}

\newcommand{\atmodel}{the analog transport model}
\newcommand{\R}{\mathbb{R}}
\newcommand{\E}{\mathbb{E}}
\newcommand{\dd}{\mathrm{d}}

\providecommand{\doi}[1]{\url{https://doi.org/#1}}

\AtBeginDocument{\nolinenumbers\let\internallinenumbers\relax}
\newcommand{\tablefont}{\fontsize{10.5}{12.5}\selectfont}
\newcolumntype{L}[1]{>{\raggedright\arraybackslash}p{#1}}
\newcolumntype{Y}{>{\raggedright\arraybackslash}X}
\hypersetup{pdftitle={Efficient Analog-Initialized Latent Transport for Spatially Coherent Probabilistic Downscaling from Global to Kilometer Scales},pdfauthor={Ophelia Miralles}}
\title{Efficient Analog-Initialized Latent Transport for Spatially Coherent \\
Probabilistic Downscaling from Global to Kilometer Scales}
\authors{Oph{\'e}lia Miralles\correspondingauthor{Oph{\'e}lia Miralles,
Norwegian Meteorological Institute, Oslo, Norway; ophelia.miralles@met.no}}
\affiliation{Norwegian Meteorological Institute, Oslo, Norway}
\abstract{Probabilistic downscaling must represent kilometer-scale structure left unresolved by a deterministic regional prediction. We introduce an analog-initialized latent transport method that uses historical regional residuals as a meteorologically informed empirical source. A frozen graph neural network predicts the deterministic state from the Integrated Forecasting System (IFS). For each input, residuals from 20 training dates with similar IFS conditions initialize the ensemble. A conditional flow-matching model transports these fields in a carefully designed 65-factor, 100,000-node latent space, and a grid decoder reconstructs 81 atmospheric channels including precipitation at 2.5-km spacing. On a season- and cycle-balanced 2025 test, ensemble CRPS improves over deterministic mean absolute error for all 81 variables. Matched analogs improve point and spatial errors over Gaussian initialization and reduce spectral error compared with covariance matching. Comparisons with released CorrDiff, a local CorrDiff-mini model and full grid EDM provide context on the 2023 validation sample. Total training of the downscaling model takes approximately 12~h on one NVIDIA H200, and inference takes about 12~s per 81-field member. The frozen networks also generate fields over France when supplied with local static data and a high-resolution IFS-derived analog bank. The method combines empirical initialization with whole-domain latent transport for computationally practical regional ensembles.}
\begin{document}
\maketitle
\statement
Public weather services, aviation and renewable-energy forecasting need affordable kilometer-scale ensembles of temperature, humidity and wind throughout the atmosphere, alongside surface weather and precipitation. The proposed method starts each member from an analysis-derived historical residual and adjusts it to the current global weather conditions. It generates an ensemble of 81 atmospheric fields including precipitation over the regional domain. Evaluation on held-out 2025 cases shows improved probabilistic scores, with remaining errors in rainfall variability and humidity extremes. A local analog bank and static fields also support execution over a new region with the trained transport and decoder weights fixed.
\section{Introduction}
Global numerical weather prediction describes the synoptic-scale state well,
but cannot explicitly resolve all of the local structures required by
regional applications. Coastlines,
mountains and land cover organize local wind, temperature and precipitation,
and several fine-scale states can be compatible with the same coarse forecast. Downscaling provides a way to recover regional detail, but information from the large-scale state does not
uniquely determine it: several kilometer-scale realizations can be compatible
with the same forecast. A complete downscaling model should thus retain the
large-scale forecast, add locally plausible detail and represent the remaining
uncertainty as an ensemble, with consistent behaviour across variables and
pressure levels. For regional applications, the predictive distribution must also describe how
uncertainty varies jointly across locations and atmospheric variables.
Statistical postprocessing commonly uses numerical model output and observations
to estimate and update predictive distributions at individual locations
\citep{ProbabilisticForecastsofPrecipitationinTermsofQuantilesUsingNWPModelOutput,
UpdatingShortTermProbabilisticWeatherForecastsofContinuousVariablesUsingRecentObservations}.

We represent the regional state as a deterministic downscaled field and a stochastic residual. The deterministic component captures information constrained by the global atmospheric state and high-resolution geography. We estimate it with a graph neural network trained in the Anemoi framework \citep{anemoi2024}, which maps global fields and regional static predictors onto the native regional grid. We define the residual as the difference between this prediction and the regional analysis. It contains kilometer-scale variability left unresolved by the finite deterministic model, including both conditional variability and errors in its prediction.

Graph neural networks have previously been used for regional forecasting and nowcasting \citep{nipen2024regional,smith2026bridgingshortmediumrangeweather,miralles2026ainca,miralles2026pointwisepointlessmultimodalablation}. Spectral objectives and probabilistic scores can reduce the smoothing of uncertain small-scale features \citep{nordhagen2025,lang2024aifscrps,taggart2026equitablescoringfunctionsoptimal}. In many forecasting applications, however, the input already contains a high-resolution regional state from which local structures can evolve. Global-to-regional downscaling provides no regional state at the requested valid time. The stochastic model must thus generate the unresolved structures directly from the coarse global state and regional geography.

Using generative deep learning models is a way to address this missing information. Adversarial models can recover local variability in wind and precipitation when provided with
topography \citep{miralles2022downscaling,leinonen2020stochastic}, but adversarial losses provide a weak indicator of forecast quality and can make training and model selection difficult. Elucidated Diffusion Models (EDM) have shown very promising results for global-to-regional downscaling in Taiwan \citep{mardani2025corrdiff}, and for
large-scale climate downscaling \citep{lebrazidec2026anemoiD2}.

Standard EDM formulations corrupt the target with Gaussian noise and learn the corresponding reverse process. Our regional state combines variables with very different distributions: precipitation is zero-inflated and heavy-tailed, humidity is constrained by saturation, and the remaining atmospheric residuals span several vertical levels and spatial scales. EDM can approximate such distributions, but a single Gaussian corruption process does not encode their physical support. The model must learn these constraints while also constructing coherent multivariate spatial structure from noise. Applying EDM directly to the complete three-dimensional state on a million-point regional grid would also be computationally expensive. We instead propose an analog-initialized latent transport model, in which the generative model is trained in a compressed space and starts from a coherent spatial and cross-variable distribution.

The key idea behind the proposed model is to start generation from an analysis-derived historical weather pattern. We call a historical state an \emph{analog} when its global atmospheric state resembles the current one. Conventional analog ensembles use retrieved historical outcomes directly as forecasts \citep{dellemonache2013analog}, while retrieval-augmented generators generally use them as conditioning information or external memory \citep{blattmann2022retrieval,sheynin2023knn}. Here, each retrieved regional residual initializes one ensemble member, and a learned transport model adjusts it to the current large-scale weather and regional geography. We retain the 64 nearest historical states as a candidate pool and sample 20 distinct analogs from it. The neighbourhood restricts the source fields to broadly similar synoptic situations, while sampling provides the ensemble with distinct starting states. Analog retrieval therefore defines the initial distribution supplied to the generative model.

The second important aspect is efficiency: we run the expensive generative step in a deliberately
designed latent space. Principal components reduce 80 jointly generated
variables to 65 factors, and a 100,000-node regional mesh reduces the spatial
dimension while preserving whole-domain connectivity. The resulting latent is
12.5 times smaller than the native 80-channel grid. A velocity model trained by
flow matching then moves each historical residual toward the current
conditional distribution. Stochastic interpolants provide the
formal basis for learning this transport between non-Gaussian, paired endpoint
distributions \citep{albergo2025stochastic,albergo2024couplings}. Finally, a
latent-to-grid decoder reconstructs all atmospheric fields over the complete
regional domain. Precipitation follows a separate transformed path because of
its strongly non-Gaussian distribution.

Our contribution is, therefore, an efficient probabilistic downscaling system that joins
these two design choices: physically informed analog initialization and
full-domain latent transport. The system generates 80 atmospheric channels
through 12 pressure levels together with precipitation on a 2.5-km regional
grid. We test whether changes to the historical source bank remain visible
after transport and decoding, assess pointwise calibration and spatial scale
content, and compare common surface fields with two full-grid
diffusion systems. We also report the selected-system training and inference cost
and run the same frozen transport and decoder over France using an IFS-derived historical
source bank and local static data. Together, these experiments test whether a
realistic empirical starting distribution can make broad regional ensemble
generation both spatially credible and computationally accessible.
\section{Methods and data}
\label{sec:methods}
\subsection{Method overview}
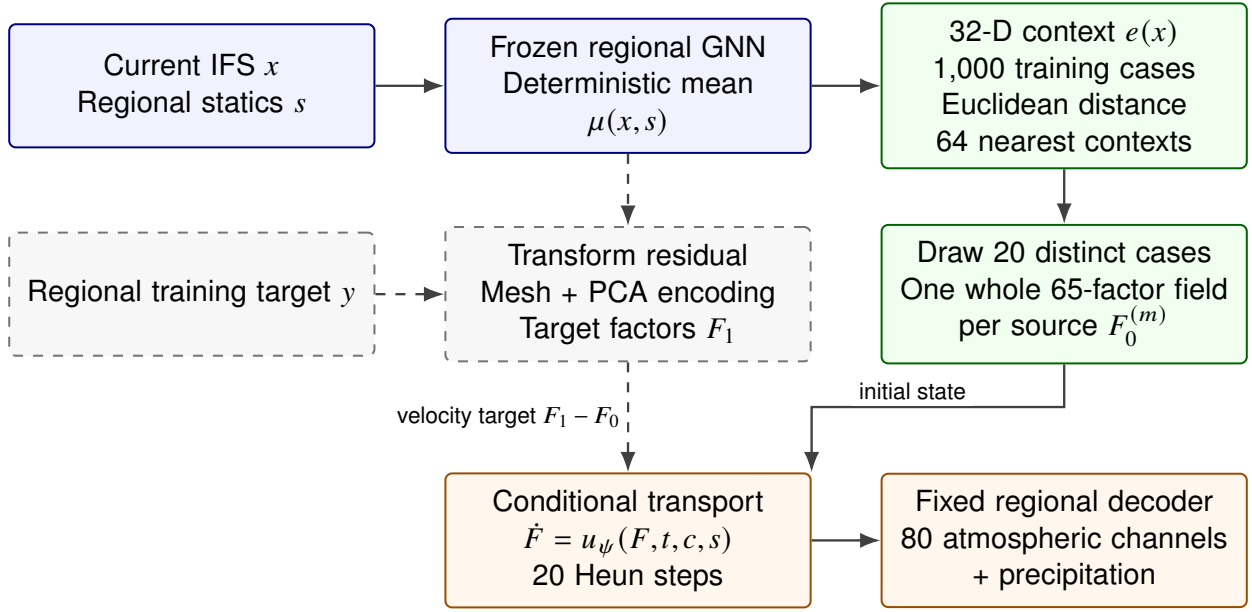
\begin{figure}[t]
\centering
\resizebox{\linewidth}{!}{\begin{tikzpicture}[font=\sffamily\small,>=Latex,
 box/.style={draw=black!55,rounded corners=2pt,align=center,text width=35mm,minimum height=13mm,inner sep=5pt,line width=.6pt},
 data/.style={box,fill=blue!5,draw=blue!50!black},
 source/.style={box,fill=green!6,draw=green!40!black},
 model/.style={box,fill=orange!7,draw=orange!60!black},
 train/.style={box,dashed,draw=black!55,fill=black!3},
 ar/.style={->,line width=.7pt,draw=black!75},
 tr/.style={ar,dashed},
 label/.style={font=\sffamily\scriptsize,fill=white,inner sep=2pt}]
\node[data] (ifs) at (0,3.6) {Current IFS $x$\\Regional statics $s$};
\node[data] (det) at (4.6,3.6) {Frozen regional GNN\\Deterministic mean\\$\mu(x,s)$};
\node[source] (search) at (9.2,3.6) {32-D context $e(x)$\\1,000 training cases\\Euclidean distance\\64 nearest contexts};
\node[train] (y) at (0,1.4) {Regional training target $y$};
\node[train] (f1) at (4.6,1.4) {Transform residual\\Mesh + PCA encoding\\Target factors $F_1$};
\node[source] (f0) at (9.2,1.4) {Draw 20 distinct cases\\One whole 65-factor field\\per source $F_0^{(m)}$};
\node[model] (flow) at (4.6,-1.2) {Conditional transport\\$\dot F=u_\psi(F,t,c,s)$\\20 Heun steps};
\node[model] (decoder) at (9.2,-1.2) {Fixed regional decoder\\80 atmospheric channels\\+ precipitation};
\draw[ar] (ifs) -- (det);
\draw[ar] (det) -- (search);
\draw[ar] (search) -- (f0);
\draw[tr] (y) -- (f1);
\draw[tr] (det) -- (f1);
\draw[tr] (f1) -- node[label,left] {velocity target $F_1-F_0$} (flow);
\draw[ar] (f0.south) -- ++(0,-.45) -| node[label,above,pos=.3] {initial state} (flow.north east);
\draw[ar] (flow) -- (decoder);
\node[font=\sffamily\scriptsize,anchor=north,align=center] at (4.6,-2.65)
{};
\end{tikzpicture}}
\caption{The complete analog-initialized latent transport system. At
inference (solid arrows), the current N320 IFS state and regional static data
first produce a deterministic downscaled field. A 32-component description
of the same IFS state is used to find similar training dates and draw 20
distinct analog residuals. Each residual is evolved in the 65-factor,
100,000-node latent space using 20 Heun updates, then the latent-to-grid
decoder reconstructs and adds the stochastic correction. Dashed arrows show regional targets used to form training residuals and fit the transport and decoder.
Precipitation follows the same logic through a separate transformed path. The diagram uses $\mu(x,s)$ for $D_\theta(x,s)$ and $c$ for the conditioning features $c(x)$.}
\label{fig:workflow}
\end{figure}

Let $x$ be the global analysis at the requested valid time, $s$ the
high-resolution static fields for the regional domain and $y$ the verifying
regional state. The complete
inference chain has four steps (Fig.~\ref{fig:workflow}). First, a frozen
deterministic downscaler $D_\theta(x,s)$ predicts the part of the regional
state that is supported by the global input and local geography. Second, the
current global state is compared with archived global states, and 20 past dates are
selected from a neighbourhood of similar situations. The stored regional
residual from each selected date is called an analog and initializes one ensemble
member. Third, a latent transport model transports each analog toward the
residual distribution appropriate for the current global state. Fourth, a frozen
latent-to-grid decoder $G_\phi$ reconstructs the residual on the regional grid
and adds it to $D_\theta(x,s)$.

The three neural components have distinct roles: $D_\theta$ supplies the deterministic downscaled state, the transport generates residual variability, and $G_\phi$ reconstructs the regional fields on the high-resolution grid. The latent representation is designed to evolve the full atmospheric state over the connected domain.

\subsection{Regional data and evaluation split}
Global inputs are operational analyses from the European Centre for Medium-Range Weather Forecasts (ECMWF) Integrated Forecasting System (IFS). Unless another resolution is stated explicitly, ``IFS'' means
the N320 reduced-Gaussian representation throughout this paper.
The primary regional target $y$ is the Nordic MEPS system, based on
HARMONIE--AROME \citep{muller2017,frogner2019a}, on a
$1069\times949$ grid with $1{,}014{,}481$ points at approximately 2.5-km
spacing. We estimate the conditional distribution $p(y\mid x,s)$ for co-valid analyses; precipitation is the associated six-hour accumulation. The experiments assess analysis downscaling at matched valid times.

The training period extends from 5 February 2020 to 30 September 2023. The validation set contains 50 cases from October to December 2023 and is used to inform archive size and retrieval neighbourhood. The test period comprises 96 cases from 1 January to 3 October 2025, selected from IFS availability. The sample contains 24 cases per meteorological seasonal stratum, with six cases at each of the 0000, 0600, 1200 and 1800 UTC cycles within each stratum. Winter covers January--February and autumn covers September--3 October. Training, validation and test dates are disjoint.

Presented results for the selected model use the test set unless stated otherwise. The
evaluation covers 81 variables: geopotential, temperature, horizontal and
vertical wind components, and specific humidity at 50, 100, 150, 200, 250,
300, 400, 500, 700, 850, 925 and 1000 hPa; 2-m temperature and dew point,
10-m wind components, surface and mean sea level pressure, skin temperature
and total column water; and six-hour accumulated precipitation in
millimetres.

The 80 non-precipitation variables share one transport model. Precipitation
uses a separate transport because its distribution contains a point mass at
zero and a long positive tail, which destabilized the PCA fit (see Section~\ref{sec:res}). The joint decoder returns all 81 fields on the same
regional grid; humidity and precipitation use the transformations specified in Appendix~B. 

\subsection{Residual target and spatial--factor latent space}
\label{sec:res}
The deterministic downscaler $D_\theta$ is a frozen graph neural network. For
each channel $v$, its error relative to the regional analysis defines the
standardized residual
\begin{equation}
r_v=\frac{T_v(y_v)-T_v\{D_{\theta,v}(x,s)\}}{a_v},
\label{eq:residual}
\end{equation}
where the variable transform $T_v$ and scale $a_v$ are estimated from the
training period. 

\begin{figure}[t]
\centering
\includegraphics[width=\linewidth]{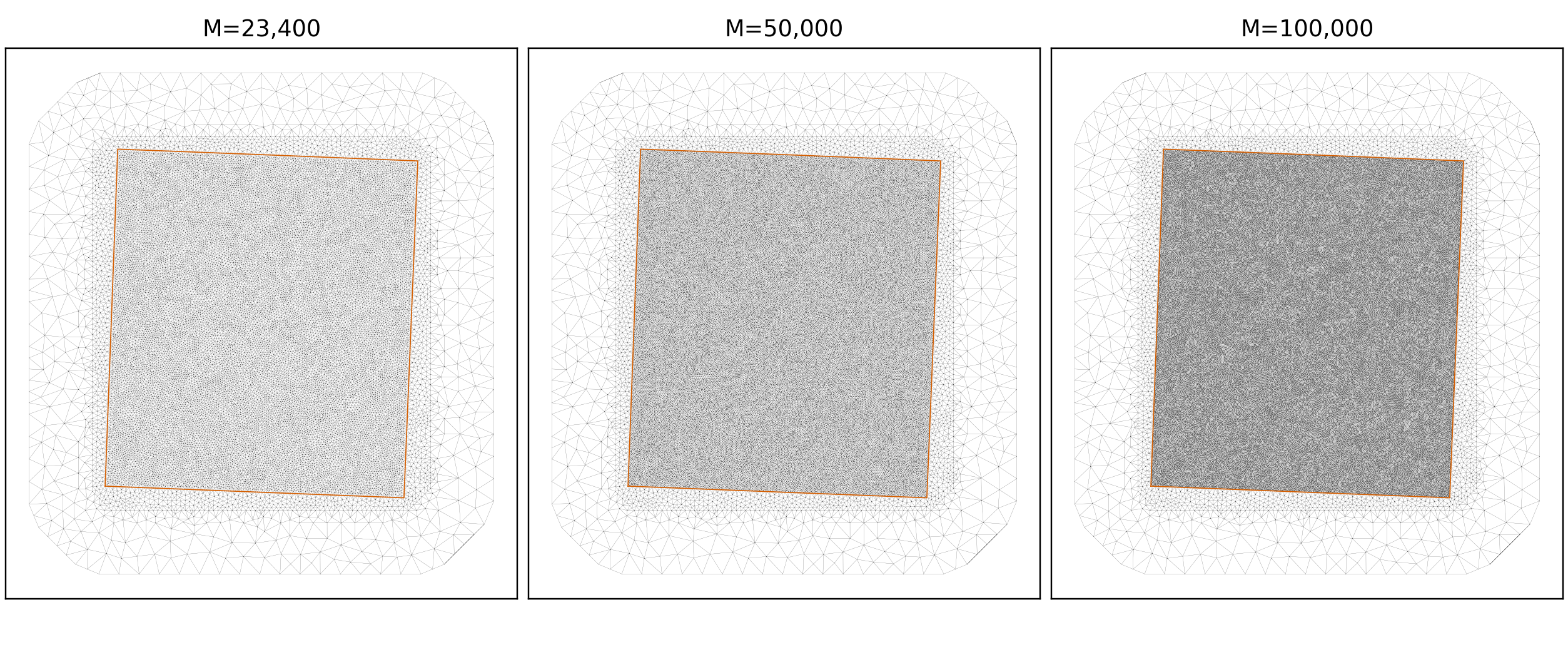}
\caption{Triangular meshes with a progressively increasing number of nodes from left to right, covering the Nordic regional
domain. Nested refinement inserts centroids of the largest interior
Delaunay triangles.}
\label{fig:mesh}
\end{figure}

We compress $r$ along both its variable and spatial dimensions. A principal-component basis
$B\in\R^{80\times65}$ maps the 80 non-precipitation variables to
$K=65$ factors. A fixed grid-to-mesh operator $E$ then represents these
factors on $M=100{,}000$ nodes spanning the complete regional domain to produce the target latent residual:
\begin{equation}
F_1=E(rB)\in\R^{M\times K},\qquad M=100{,}000,\quad K=65.
\label{eq:encoding}
\end{equation}
The 100,000-node triangular mesh is obtained by repeatedly inserting centroids of the largest interior Delaunay triangles and using barycentric operators for mesh-to-grid projection \citep{rue2017bayesian}. The mesh size was selected so that native-grid reconstruction retained a median $\approx 94\%$ of the original variance (Figure~\ref{fig:mesh}). Principal component analysis (PCA) compresses the variable dimension. Channel PCA is fitted by eigendecomposition of the empirical covariance of the standardized atmospheric residuals, pooled over 800 training dates and 5,000 spatial locations per date (approximately four million samples). We select the leading 65 components based on explained variance, residual dependence, and subspace stability computed on a separate 200-date subset. We refer to this PCA--mesh representation as the latent throughout the paper. The latent contains 6.5 million values, compared with 81.2 million values on the native 80-channel grid, which represents a 12.5-fold compression. 

\subsection{Selecting historical analogs}
An analog is the residual from a past training date whose global IFS
state is similar to the current one. The analog archive contains 1,000 pairs; the archive size was chosen by balancing model skill and storage requirements. Each archive entry $i$ corresponds to a historical training date $t_i$. It contains an IFS descriptor $e(x_i)$ and the complete latent regional residual $F_i$ computed from the paired regional analysis at that historical date:
\begin{equation}
\mathcal A=\{(e(x_i),F_i)\}_{i=1}^{1000}.
\end{equation}
For a new requested valid time $t^\ast$, retrieval compares the current IFS descriptor $e(x^\ast)$ with the stored historical descriptors $e(x_i)$. Once a historical index $i$ has been selected, its paired residual $F_i$ initializes an ensemble member. During training, the regional residual at the target date is used as the transport endpoint $F_1$, and exact descriptor self-matches are excluded from retrieval.

\begin{figure}[t]
\centering
\includegraphics[width=0.75\linewidth]{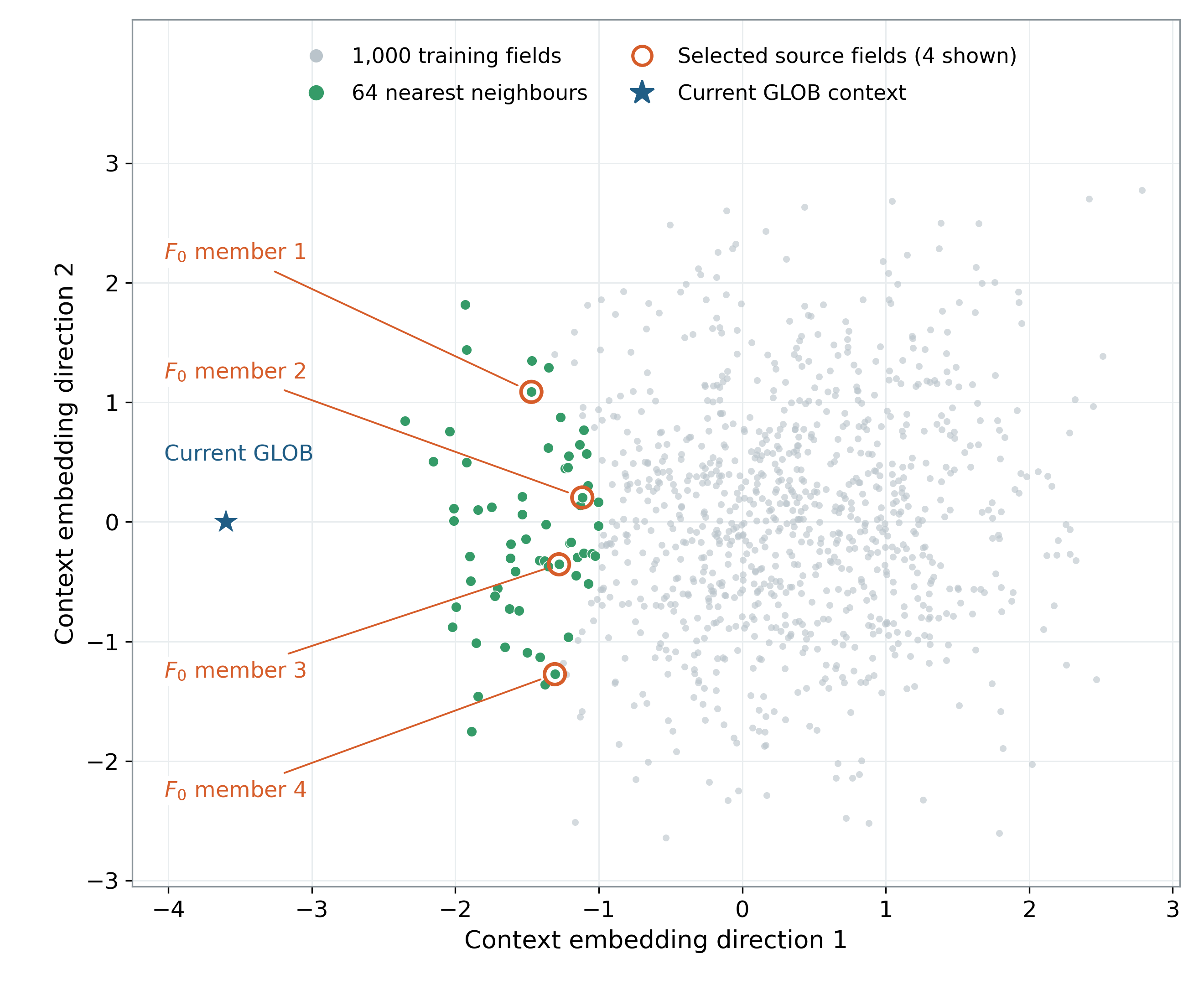}
\caption{Schematic illustration of analog selection. The current N320 IFS state (star) is compared with
1,000 archived training states (grey) using Euclidean distance in the
32-dimensional whitened similarity space. The 64 closest dates form the
candidate pool (green), and 20 different dates are drawn uniformly for the
ensemble; four draws are outlined in orange. Each draw supplies the complete
regional residual from one past date, including all 65 factors. Only two axes
are used here to visualize the 32-dimensional search.}
\label{fig:retrieval}
\end{figure}

The descriptor $e(x_i)$ is constructed from frozen deterministic-downscaler predictions for eight variables representative of the atmospheric state: 2-m temperature, 10-m zonal and meridional wind, mean sea-level pressure, 500-hPa geopotential, 850-hPa temperature, 700-hPa specific humidity and total column water. We project the eight fields to the mesh and sample 256 fixed farthest-point nodes. The resulting 2,048 values are standardized with archive-only statistics, reduced by singular-value decomposition to 32 components and whitened to zero mean and unit covariance. Each retained direction then contributes on a common scale to the retrieval distance. For the current
IFS global state $x$, the distance to an archive case $i$ is simply
\begin{equation}
d_i(x)=\lVert e(x)-e(x_i)\rVert_2.
\label{eq:retrieval}
\end{equation}
We evaluated neighbourhoods of 12, 24, 32 and 64 cases on the validation set.
Larger neighbourhoods reduced RMSE and CRPS and increased the spread--skill
ratio and spatial correlation, with gains levelling off between 32 and 64
cases. We selected the 64 cases with the smallest values of $d_i(x)$ as the
candidate neighbourhood (Fig.~\ref{fig:retrieval}).

At inference time, we draw 20 cases uniformly from these 64 candidates without
replacement. The complete $F_i$ attached to the selected date becomes the
initial latent residual $F_0$ of one member. The 64-candidate neighbourhood supplies meteorologically similar weather patterns, and drawing 20 cases gives
the ensemble 20 distinct initial states while keeping inference affordable and
matching the evaluation ensemble size.

Thus, after the current IFS input is fixed, the random selection of an analog
is the source of ensemble diversity; the subsequent trajectory is
deterministic. We train and apply the generative model entirely in the PCA--mesh latent space. The transport loss is evaluated on latent residuals, and the generated trajectory is integrated in the same space before being decoded to the regional grid. We therefore refer to this stage as analog-initialized latent transport.

\subsection{Analog-initialized latent transport}
The generative component is a conditional probability-flow model trained by
flow matching. Its empirical source distribution is transported to the conditional residual distribution by a learned velocity field. 
For each
training target $F_1$, we select an analog $F_0$ using only the corresponding
IFS input. At a randomly drawn path time $t\sim\mathcal U(0,1)$, the training
state and its desired velocity are
\begin{equation}
F_t=(1-t)F_0+tF_1,\qquad v^\star=F_1-F_0.
\end{equation}
The velocity network $u_\psi$ is trained to infer this displacement from the
intermediate latent state, path time and current-weather information:
\begin{equation}
\mathcal L(\psi)=\E\left[
\lVert u_\psi(F_t,t,c(x))-(F_1-F_0)\rVert_W^2\right],
\label{eq:loss}
\end{equation}
where $\lVert A\rVert_W^2=\sum_{m,k}w_{mk}A_{mk}^2$ is the weighted velocity mean-square loss. The conditioning $c(x)$ is the frozen deterministic regional prediction projected to the latent mesh. Regional context $s$ enters through this deterministic prediction. The 32-component descriptor $e(x)$ ranks archive cases, and the full mesh representation $c(x)$ conditions the velocity field. 

At inference, the selected analog is the initial condition of the learned
ordinary differential equation
\begin{equation}
\frac{\dd F}{\dd t}=u_\psi(F,t,c(x)),\qquad F(0)=F_0.
\label{eq:ode}
\end{equation}
We then use Heun's method for sampling from Equation~\ref{eq:ode}. Heun's method is a second-order predictor--corrector integrator, in which each update predicts the next latent state and then corrects it using a
second velocity evaluation. 

The deterministic downscaler, velocity network and high-resolution decoder all follow the encoder--processor--decoder graph architecture
used in data-driven weather models \citep{lam2022graphcast,lang2024aifs} and are
implemented in Anemoi \citep{anemoi2024}. The velocity network uses graph connections at 50,
150 and 350 km, eight graph-transformer layers, 512 hidden channels and eight
attention heads. The deterministic downscaler and the analog-conditioned latent transport contain respectively $\approx 27$M and $\approx 58$M learned parameters. Appendix~C records the training schedules and frozen stages.

\subsection{Latent-to-grid decoding and physical variables}
The latent-to-grid decoder is denoted by $G_\phi$. The full decoding process combines a fixed linear reconstruction through the
mesh-to-grid operator $P$ and PCA basis $B$ with a learned graph component
$g_\phi$:
\begin{equation}
\widehat r=G_\phi\!\left(\widehat F,x,s,D_\theta(x,s)\right)
=(P\widehat F)B^\top
+g_\phi\!\left(\widehat F,x,s,D_\theta(x,s)\right).
\label{eq:decoder}
\end{equation}

The final downscaled field is the deterministic prediction plus the generated
residual in the variable's transformed coordinate,
\begin{equation}
\widehat y_v=T_v^{-1}\!\left[
T_v\{D_{\theta,v}(x,s)\}+a_v\widehat r_v
\right].
\label{eq:decode}
\end{equation}
The learned decoder component $g_\phi$ contains $\approx 23$M parameters and is trained to reconstruct training residuals from their encoded latent
states. Its inputs include
the transported latent, current IFS state, deterministic downscaled state, elevation,
land--sea information, land cover, spatial coordinates, time and solar features. 

Most atmospheric variables are near-Gaussian and do not undergo any specific transformation. Specific
humidity is represented as a bounded fraction of saturation at the
corresponding temperature and pressure, which couples its inverse transform to
the generated temperature. Six-hour precipitation uses a continuous
log-amount coordinate (Appendix~B) and a separate latent transport. The precipitation decoder
loss term combines pointwise reconstruction with gradient and spectral terms to
retain local boundaries and scale content. Appendix~B specifies these transforms and quality-control rules. 

\section{Verification}
\subsection{Metrics}
We evaluate ensemble-mean root-mean-square error (RMSE), continuous ranked
probability score (CRPS), spread relative to ensemble-mean RMSE, and
generated-to-target variance ratios. CRPS is a proper
score for a predictive distribution \citep{gneiting2007probabilistic}. A
deterministic forecast has CRPS equal to its absolute error, so comparisons
with a deterministic distribution use mean absolute error (MAE).

Variograms and power spectra assess spatial dependence and scale content.
Structural similarity (SSIM) and log-spectral distance (LSD) provide
complementary member-wise image diagnostics. Aggregate curve errors compare
case-averaged variograms or spectra with the corresponding target curves.
These measure the ensemble's typical scale content. SSIM assesses local
feature correspondence, and member-wise LSD compares individual spectra. External
model differences are reported as paired relative error reductions. Precipitation verification includes occurrence Brier score at 0.1~mm per six hours, reliability, positive-amount quantiles, interval coverage, fractions skill score (FSS) and connected-object counts and areas. FSS is evaluated at 0.1, 1, 5 and 10~mm thresholds and 10--100~km neighbourhoods. Seasonal scores and three event strata use the upper quartile of IFS domain 95th-percentile wind speed, 2-m temperature P95--P05 contrast and six-hour precipitation. 

Uncertainty is estimated by paired resampling of complete dates, keeping all
variables, spatial points and ensemble members of a date together. All reported test, ablation and
external comparisons use 5,000 bootstrap replicates
and percentile 95\% intervals. These intervals describe variation across
dates for the evaluated checkpoints.

\subsection{External benchmarks}
For comparison with external benchmarks, we use NVIDIA's version-1 REA6 checkpoint, loaded through Earth2Studio
\citep{nvidia2026cosmo}, and refer to it as ``CorrDiff''. This checkpoint is
a direct-field diffusion transformer with an 18-step Heun sampler that maps
ERA5 to 45 variables on the approximately 6-km COSMO-REA6 grid. During this study, first-member outputs from another regional downscaler were
made available for benchmarking \citep{stoylen2026carra_cerra_ml}. This model
maps ERA5 to the CERRA \citep{cerra2024} and CARRA
\citep{c3s_carra,schyberg2020carra} domains and produces 16 surface variables
with a graph neural network implemented in Anemoi \citep{anemoi2024}. It uses
elucidated diffusion directly on the full regional grid. We refer to this
external deployment as ``full grid EDM''. \citet{lebrazidec2026anemoiD2} describe
the model family in a separate experiment that downscales global fields from
approximately 100-km to 30-km resolution. 
We additionally evaluate a local eight-variable CorrDiff-mini model trained on the same IFS/MEPS data as our own model. Full grid EDM supplies only one archived member and partial coverage; its RMSE is thus compared with our ensemble mean on its finite footprint. Member-wise LSD uses member 1. External-model comparisons use the validation set because of external data availability.

\subsection{Source distribution comparison}
\label{sec:source-controls}
In order to assess the added value of the analog-matched initialization, we compare four initial distributions: independent standard Gaussian values, a fitted stochastic partial differential equation (SPDE) covariance source, uniformly sampled complete historical residuals, and context-matched analog residuals. Each learned arm uses a separately trained transport model initialized from scratch. Architecture, IFS conditioning, optimizer, learning-rate schedule, initialization seed schedule, a 10,000-update training budget, frozen decoder and 20-step Heun inference budget are held fixed. The raw matched-analog model passes the retrieved residual directly to the decoder and does not involve transport. The comparison uses 20 members on a calendar-balanced 32-case subset of the 2025 test. All learned arms use the checkpoint from training seed 20260901. It measures the combined effect of the source distribution and the transport learned for that distribution.

\subsection{Application across regional domains}
\label{sec:domains}
The released CorrDiff checkpoint is trained on European COSMO-REA6, whereas
our model is trained on Nordic MEPS data. We use this asymmetry in two
directions. Over the Nordics, both products are verified against MEPS even
though CorrDiff learned a different regional analysis and native resolution.
France is inside CorrDiff's training geography but outside ours.
We therefore test whether the Nordic-trained latent transport can be instantiated on a
new domain. We translate the
regional grid, replace coordinates and high-resolution static fields, and
build a local archive of 128 historical source residuals from the frozen $D_\theta$ applied to IFS N320 over France and high-resolution IFS O1280 data representing the high-resolution reference in the absence of local analysis. 
The comparison begins from global inputs of similar nominal
resolution: N320 IFS for analog transport and ERA5 for CorrDiff. We generate 20
members from both systems for 1200 UTC 7 November 2023, and display the deterministic downscaler alongside the generated field to show the regional correction.

\section{Results}
\label{sec:results}
\subsection{Source distribution ablation}
\label{sec:ablation-results}
\begin{figure}[t]
\centering
\includegraphics[width=\linewidth]{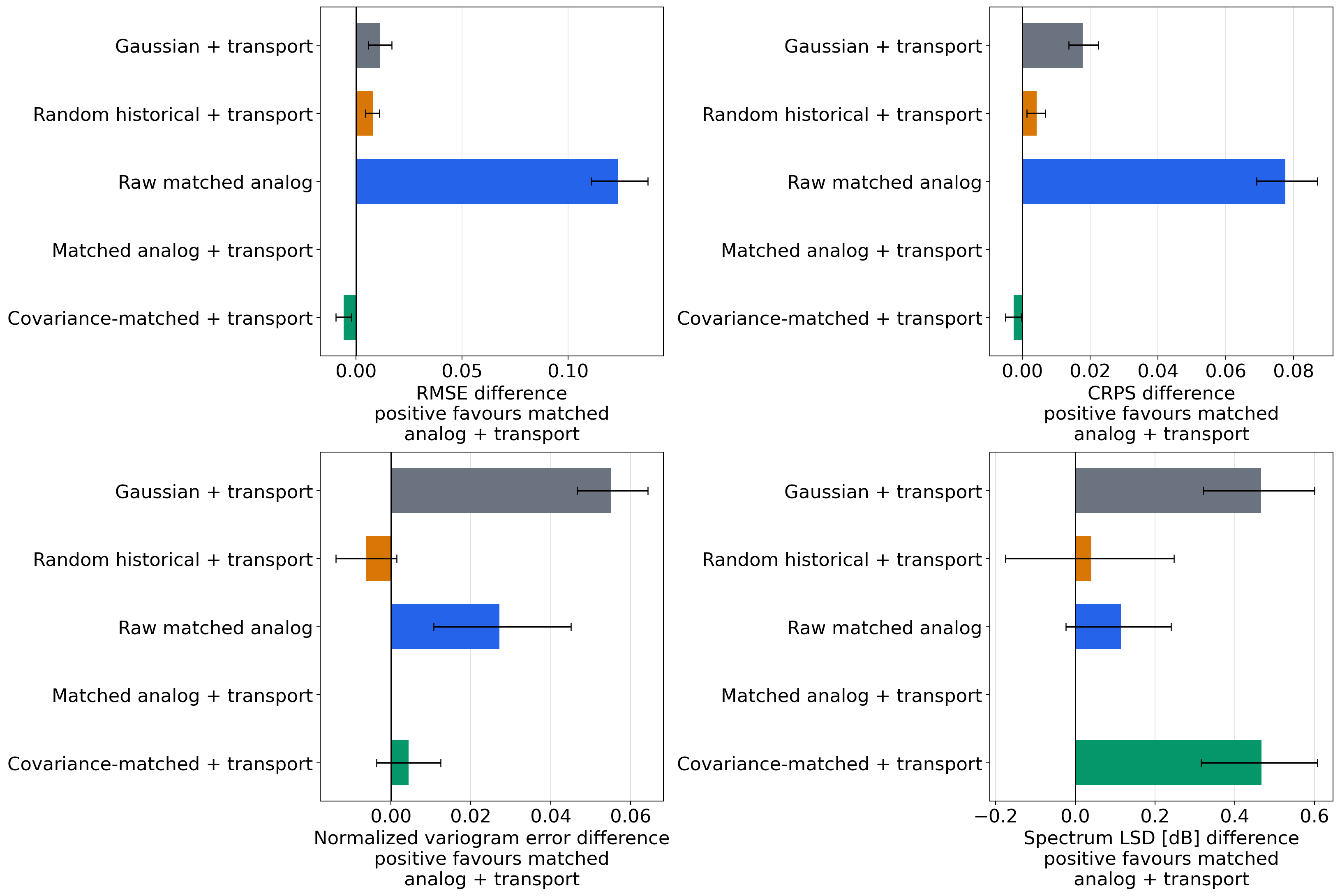}
\caption{Source distribution ablation results on 32 test cases with 20 members. Bars are control error minus matched-analog-transport error; positive values favour matched analog transport.}
\label{fig:source-ablation}
\end{figure}

Figure~\ref{fig:source-ablation} compares transport models trained with
different source distributions, alongside the raw matched-analog control
(Section~\ref{sec:source-controls}). Matched analog transport improves
RMSE, CRPS, variogram error and spectral fidelity over Gaussian
initialization. It also improves RMSE and CRPS over random historical
initialization. Applying transport to the matched analogs improves their
RMSE, CRPS and variogram error, showing that the learned adjustment adds
skill to the retrieved fields. The paired confidence intervals support
each of these improvements.
Covariance-matched initialization achieves lower RMSE and CRPS than matched
analog transport. Matched analogs better reproduce the target spectra,
reducing LSD by approximately 0.47~dB. The choice of source distribution
therefore involves a trade-off between pointwise and probabilistic accuracy
and the spatial scale content of individual members.

\subsection{Whole-domain generation recovers spatial variability}
\begin{figure}[t]
\centering
\includegraphics[width=.49\linewidth,height=0.77\textheight,keepaspectratio]{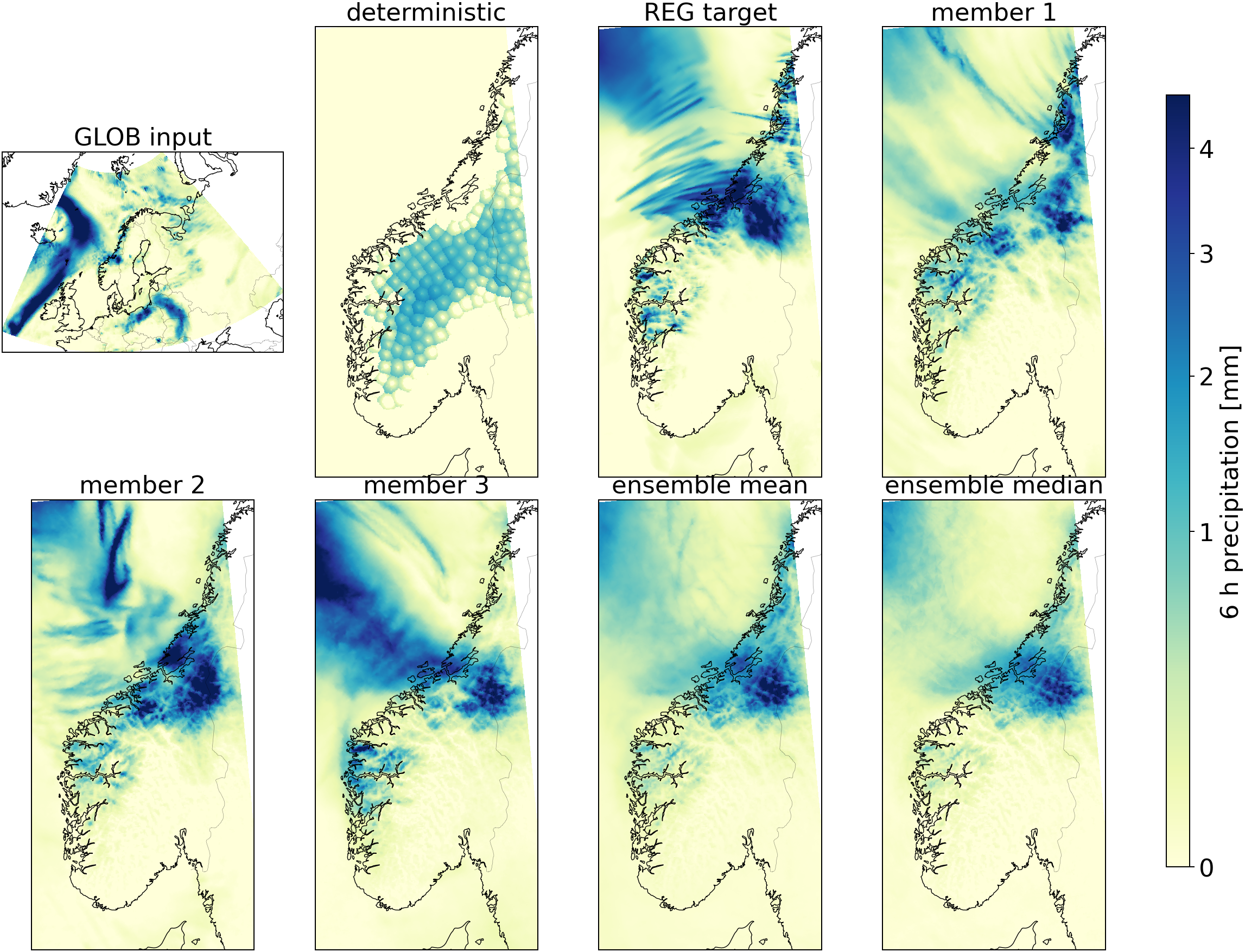}
\includegraphics[width=.49\linewidth,height=0.77\textheight,keepaspectratio]{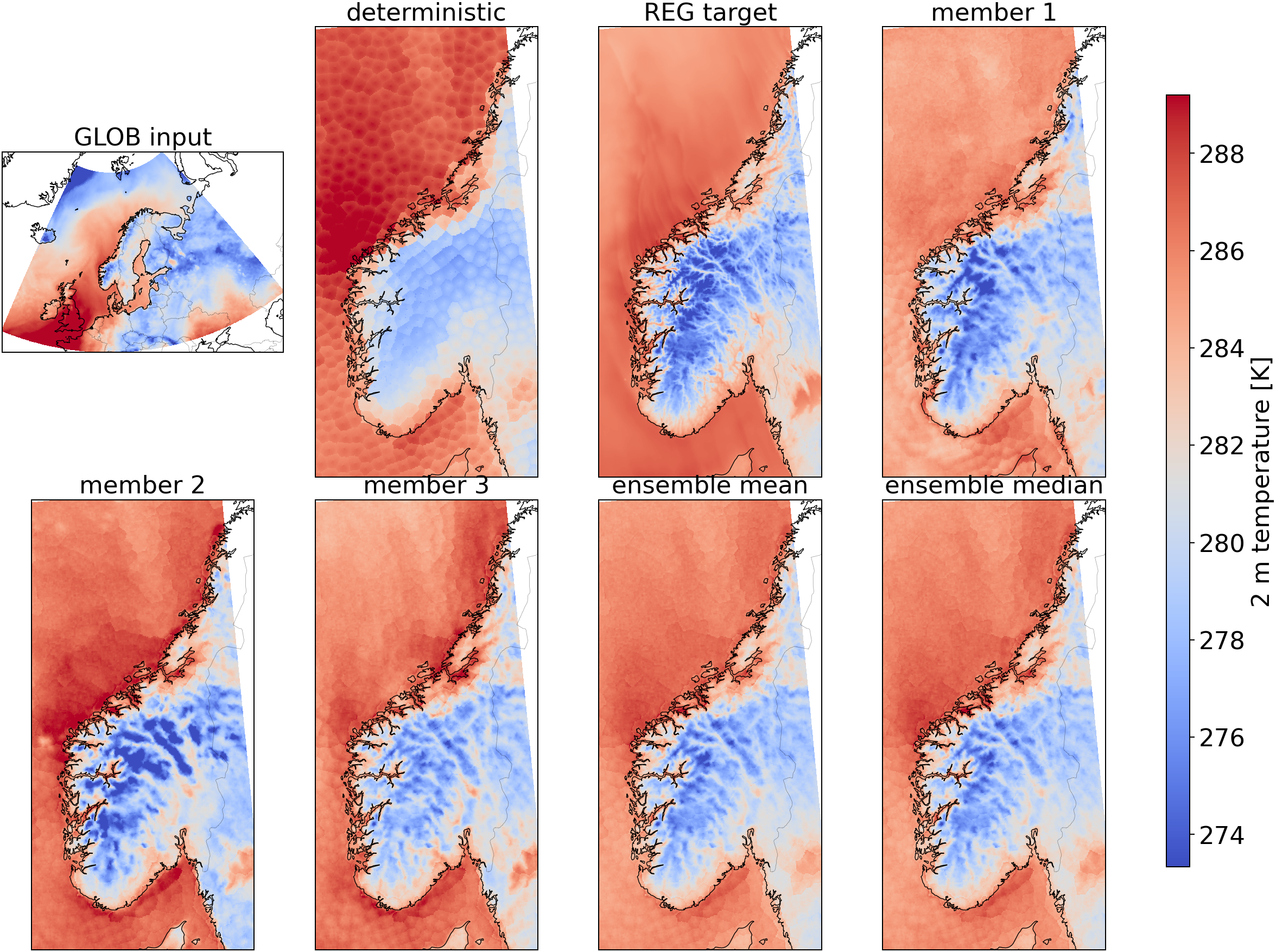}
\caption{Six-hour precipitation at 1800 UTC 1 March 2025, in millimetres (left) and surface temperature in Kelvin on 3 October 2025 (right).}
\label{fig:regional-fields}
\end{figure}

The model generates each complete $1069\times949$ regional state in one calculation. Figure~\ref{fig:regional-fields} displays a fixed native-resolution precipitation window from 1 March 2025, with the deterministic prediction, target and three named members. The members form connected rain bands and differ in their local placement and intensity. The ensemble median smooths these differences. Fine mesh-related texture remains visible in some temperature and wind samples, particularly over sea.

We assess surface-wind structure through divergence and vorticity, which
describe horizontal convergence or expansion and local rotation,
respectively. For each diagnostic, we compare its root-mean-square (RMS)
magnitude with that of the MEPS target; a ratio of one indicates matching
amplitudes. Member 1 increases the divergence RMS from about 43\% to 54\% of the
target amplitude and the vorticity RMS from about 70\% to 79\%, relative
to the deterministic downscaler.
The generated winds therefore recover some of the variability smoothed by
the deterministic downscaler. 

\subsection{Humidity and precipitation}
On the held-out 2025 test, the humidity ensemble improves CRPS over the
deterministic forecast at all 12 pressure levels and more closely reproduces
vertical humidity correlations. The upper tail remains too moist at 400,
500 and 700~hPa, where the pooled 99.8th percentiles exceed the target by
approximately 40--60\%.
For six-hour precipitation, ensemble CRPS is approximately 54\% lower than
the deterministic forecast's MAE. The ensemble also improves predictions of
rain occurrence, with a lower Brier score at the 0.1-mm threshold. Its mean
predicted wet probability is close to the observed wet frequency, with a
slight underestimate.
\begin{figure}[t]
\centering
\includegraphics[width=0.35\linewidth]{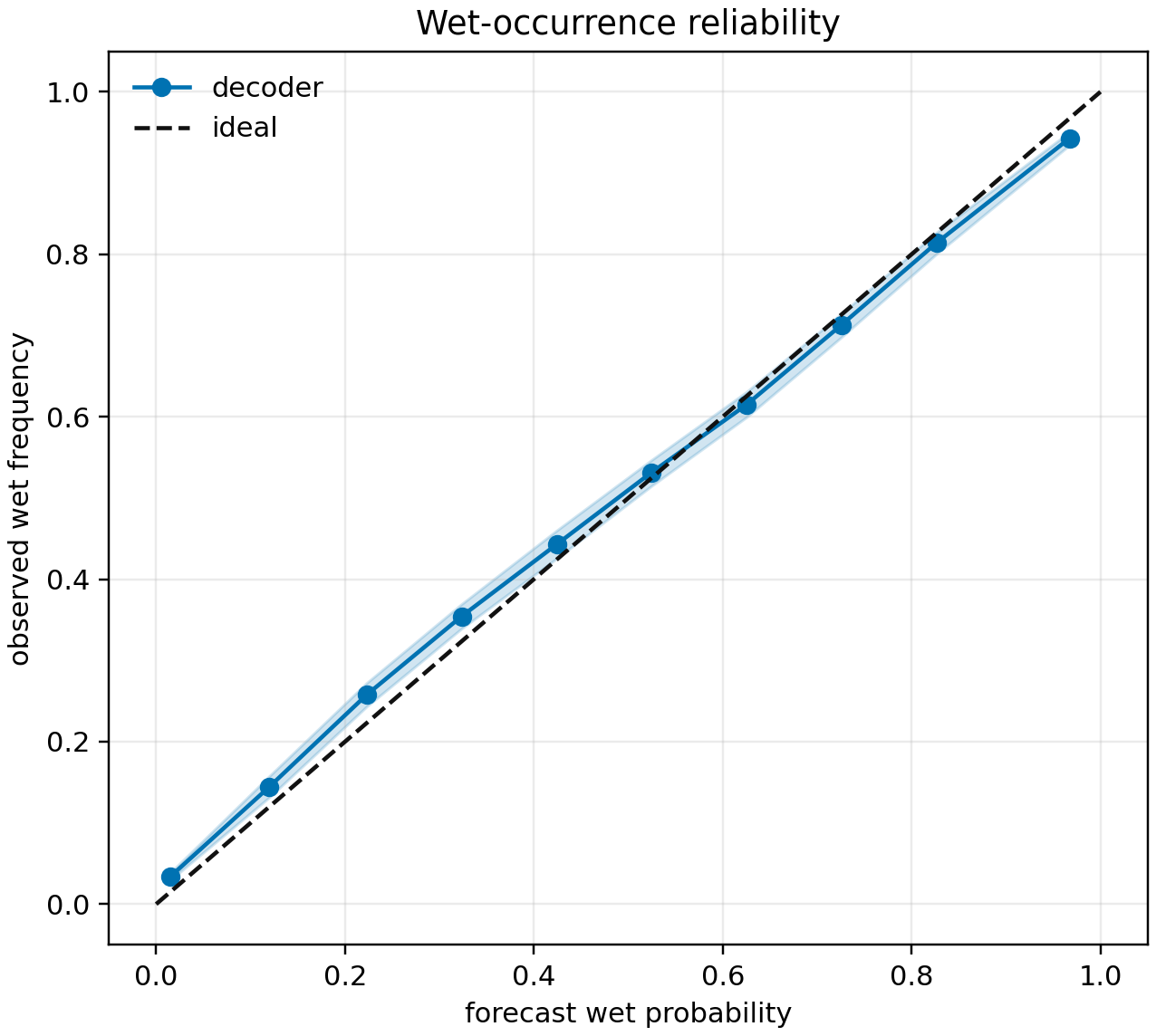}
\includegraphics[width=0.64\linewidth]{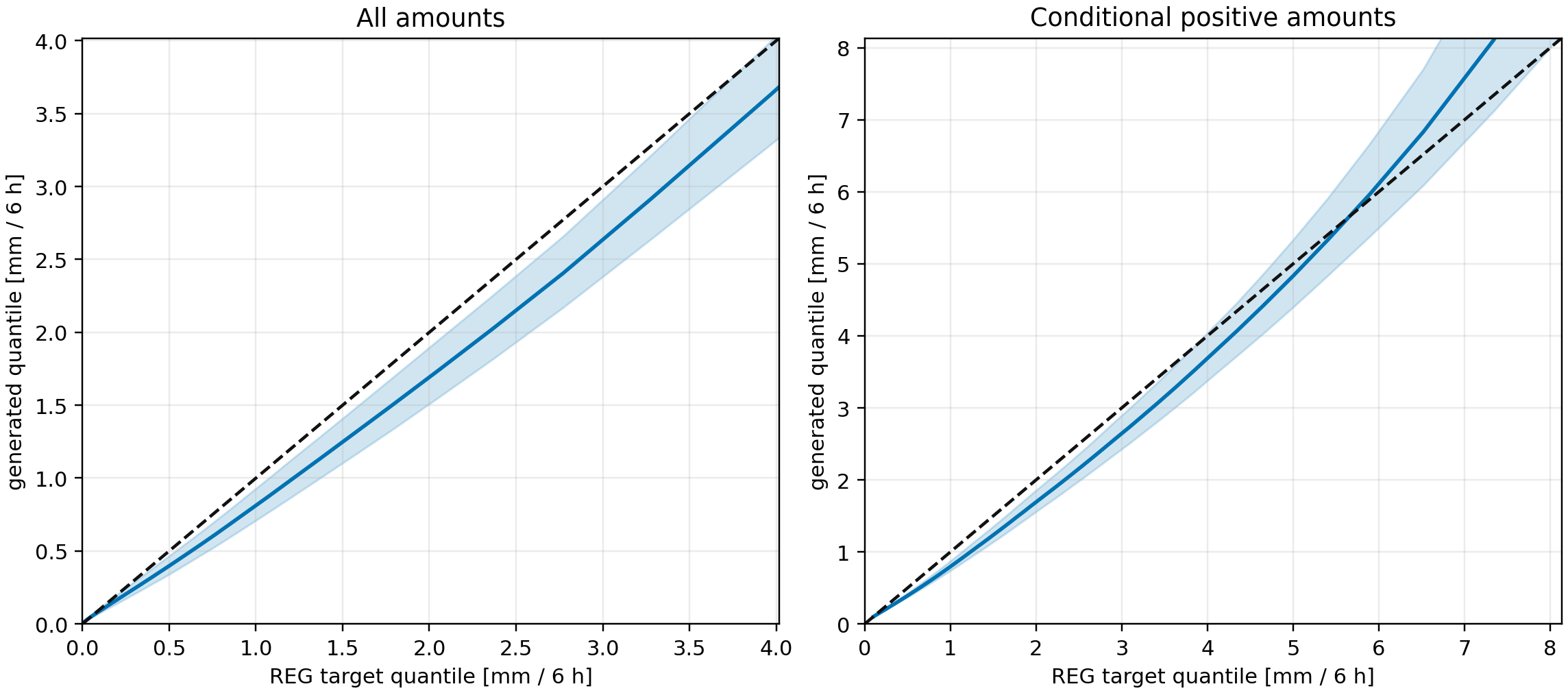}
\caption{Six-hour precipitation verification on the 96-case 2025 test. Left: occurrence reliability at 0.1~mm, comparing forecast probability with observed frequency. Right: quantile--quantile diagnostics for all amounts and conditional positive amounts. Diagonals denote agreement. Occurrence reliability and marginal amount distributions assess different aspects of the ensemble.}
\label{fig:precip-qq}
\end{figure}

\begin{figure}[t]
\centering
\includegraphics[width=0.49\linewidth]{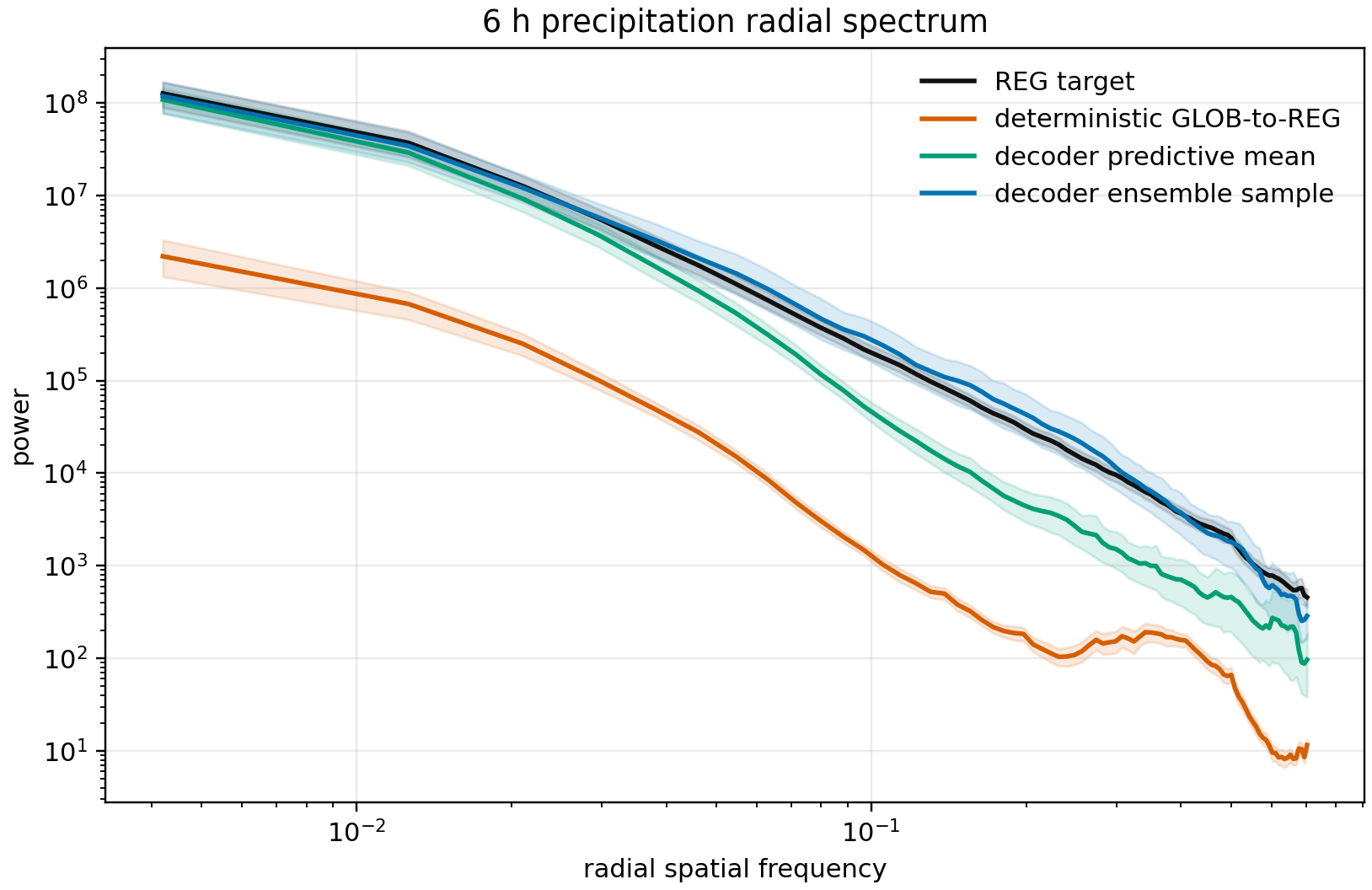}
\includegraphics[width=0.49\linewidth]{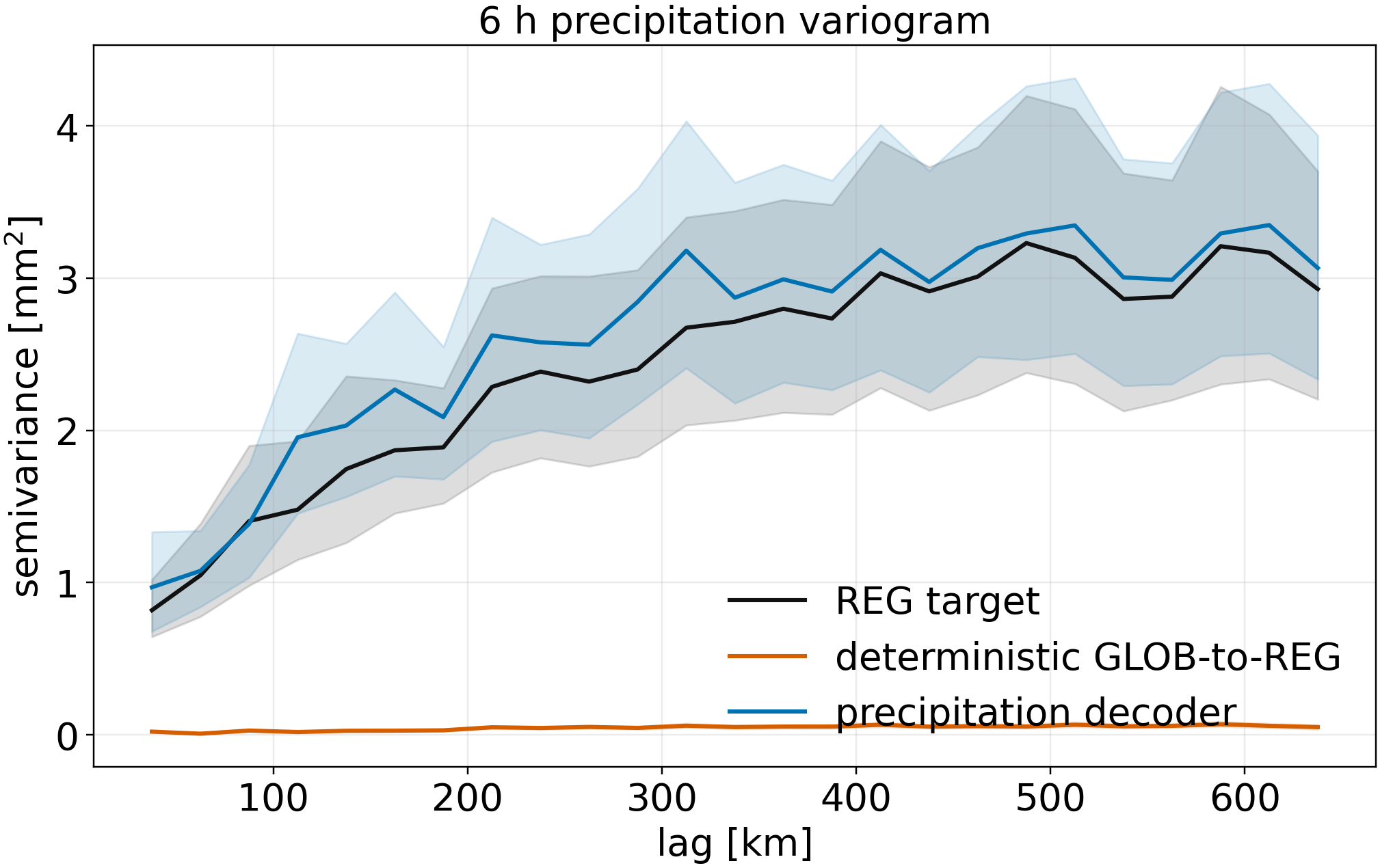}
\caption{Six-hour precipitation spectra (left) and variograms (right).}
\label{fig:precip-spatial}
\end{figure}

The precipitation ensemble has excessive overall spread and variance,
while its nominal 90\% intervals cover approximately 85\% of verifying
values. The highest quantiles of positive rainfall amounts are also
overestimated. Generated members reproduce much of the target's spatial
variability, with power spectra and variograms close to the target over
most displayed frequencies and lags (Fig.~\ref{fig:precip-spatial}).
Averaging across members smooths the finest-scale features, reducing
their power in the ensemble mean.

\subsection{Comparison with released full-grid diffusion downscalers}
\begin{figure}[t]
\centering
\includegraphics[width=\linewidth]{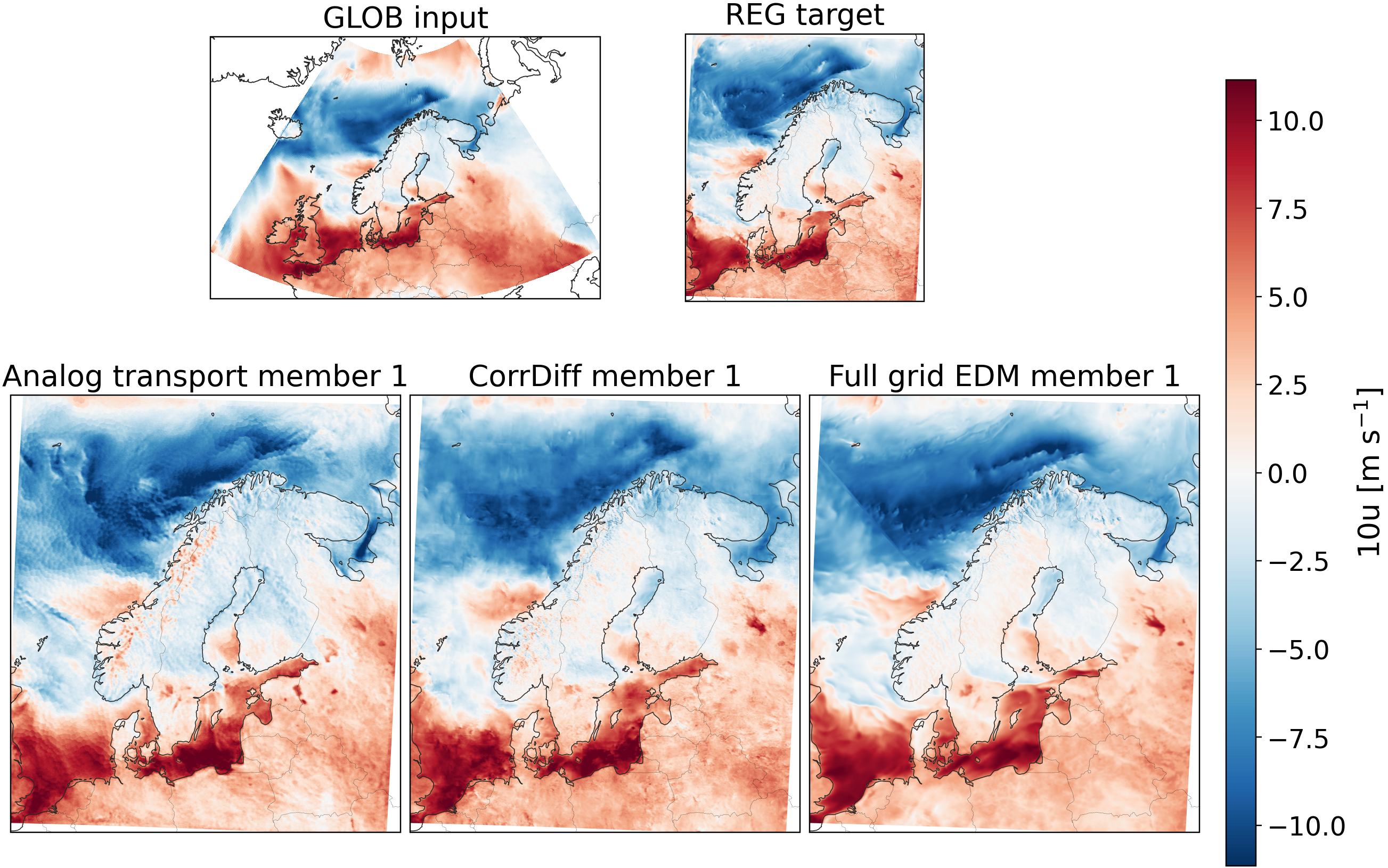}
\caption{10-m zonal wind at 1200 UTC 7 November 2023, in m s$^{-1}$.
The upper row shows the N320 IFS input and MEPS target; the lower row compares
member 1 from \atmodel{}, CorrDiff
\citep{nvidia2026cosmo} and the full grid EDM. CorrDiff is plotted
within its available coverage, with the same approximately 6-km native
output used in the quantitative comparison.}
\label{fig:wind-maps}
\end{figure}

The released checkpoints provide contextual comparisons across model
families, training domains and output scopes. On the validation set,
analog-initialized latent transport has lower single-member RMSE than CorrDiff
for 2-m temperature and both 10-m wind components
(Table~\ref{tab:corrdiff}). RMSE reductions are approximately 12--16\%, with all three confidence
intervals excluding zero. 
For 7 November 2023 at 1200 UTC, each system captures the domain-scale reversal
in zonal wind (Fig.~\ref{fig:wind-maps}).

\begin{table}[t]
\centering\tablefont
\begin{tabular}{lrr}
\toprule
Field & RMSE reduction (\%) [95\% CI] & Spectrum-error reduction (\%)\\
\midrule
2-m temperature & 16.1 [11.1, 20.8] & 34.7\\
10-m zonal wind & 12.3 [9.1, 15.3] & 42.2\\
10-m meridional wind & 14.7 [11.7, 17.6] & 44.3\\
10-m wind speed & $-1.3$ [$-4.0$, 1.4] & 36.7\\
\bottomrule
\end{tabular}
\caption{Paired relative errors of \atmodel{} against the released NVIDIA
COSMO-REA6 CorrDiff checkpoint \citep{nvidia2026cosmo}; positive values
favour \atmodel{}. Member 1 is scored on 50 October--December 2023 validation dates.}
\label{tab:corrdiff}
\end{table}

\begin{figure}[tp]
\centering
{\small\bfseries (a) 2-m temperature\par}
\includegraphics[width=\linewidth]{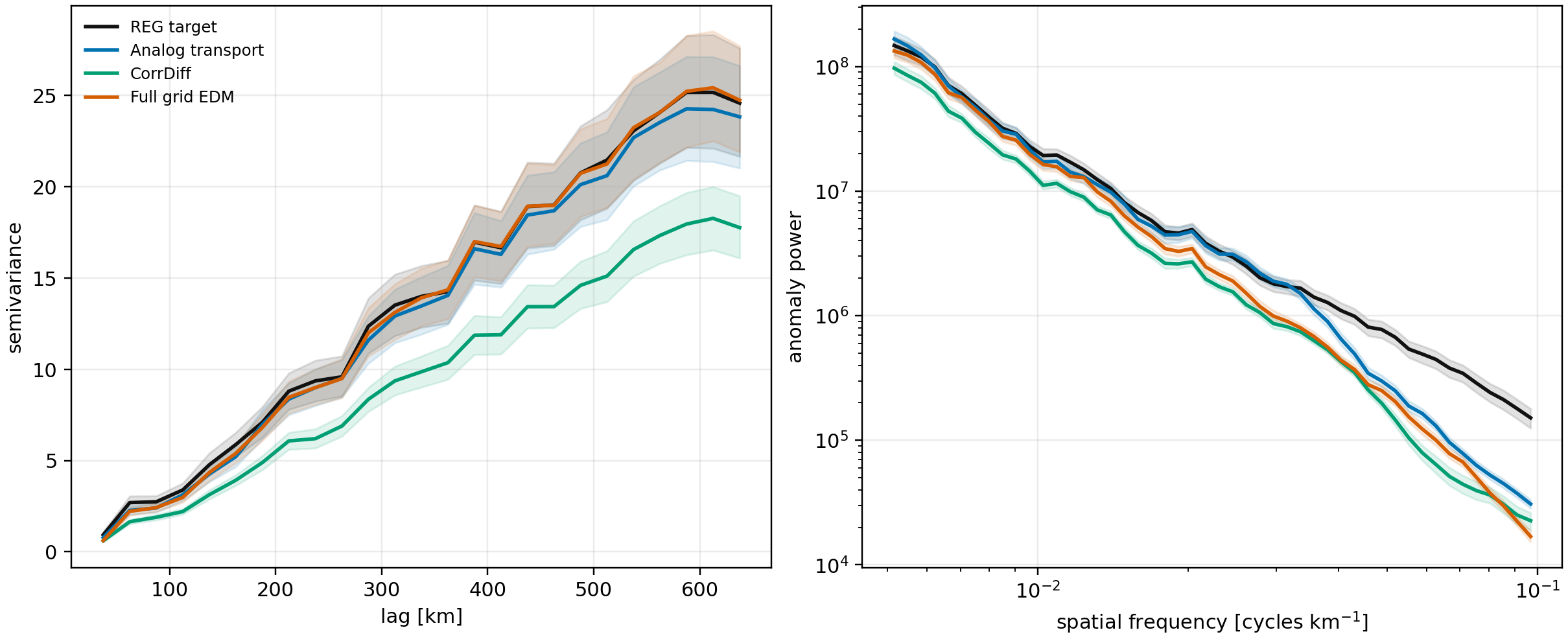}
\medskip
{\small\bfseries (b) 10-m wind speed\par}
\includegraphics[width=\linewidth]{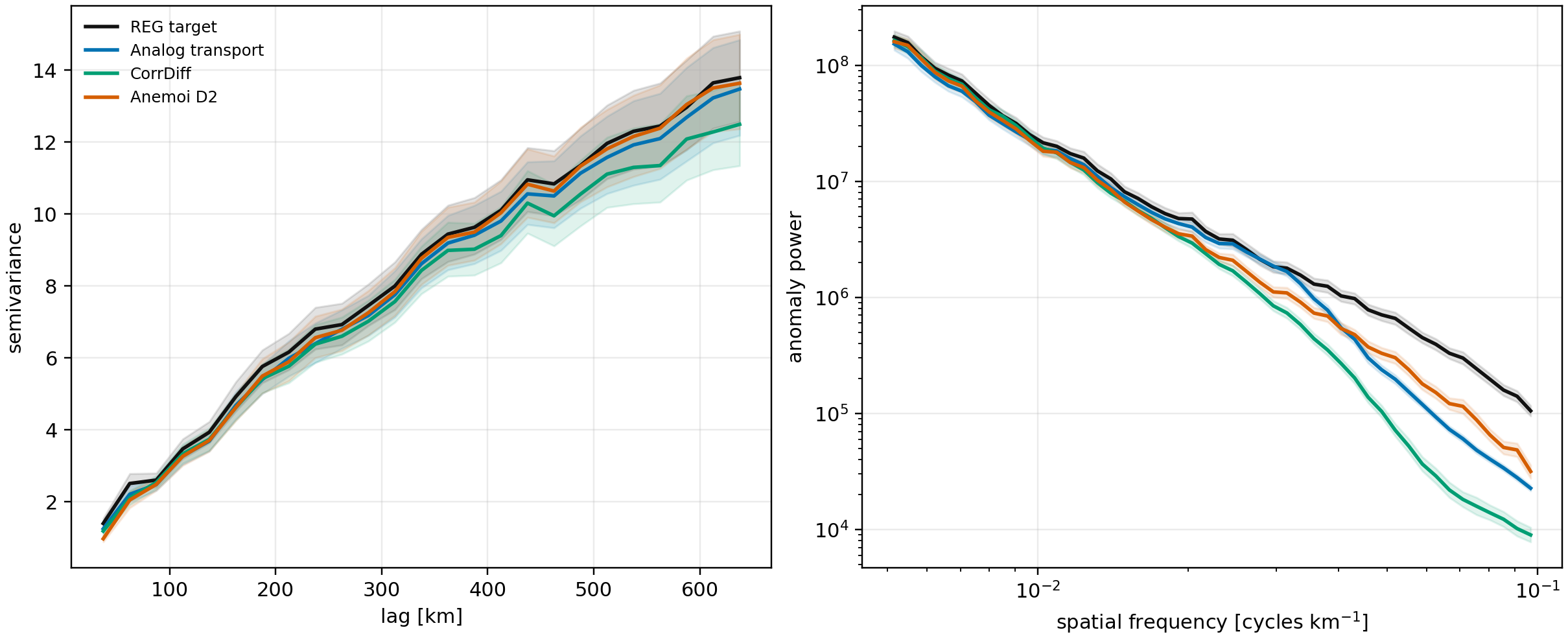}
\caption{Spatial dependence and scale content over the Nordic
domain over 50 validation dates, using member 1 from each method.
Left: aggregate variograms; right: anomaly power spectra.
Black curves denote
MEPS; blue, green and orange denote \atmodel{},
CorrDiff \citep{nvidia2026cosmo} and full grid EDM.
Shading shows 95\% case-bootstrap intervals. }
\label{fig:spatial-curves}
\end{figure}

Against CorrDiff, analog transport reduces aggregate spectrum-curve error by
approximately 35--44\% for all four shared fields. Variogram error is lower by about 87\%
for temperature and 52\% for zonal wind; the intervals for meridional wind
and wind speed include zero. CorrDiff performs better on SSIM and high-pass
SSIM, which are more sensitive to the exact location of features in an
individual member. All three
generative systems lose power at the highest frequencies (Fig.~\ref{fig:spatial-curves}).

\begin{figure}[t]
\centering
\includegraphics[width=\linewidth]{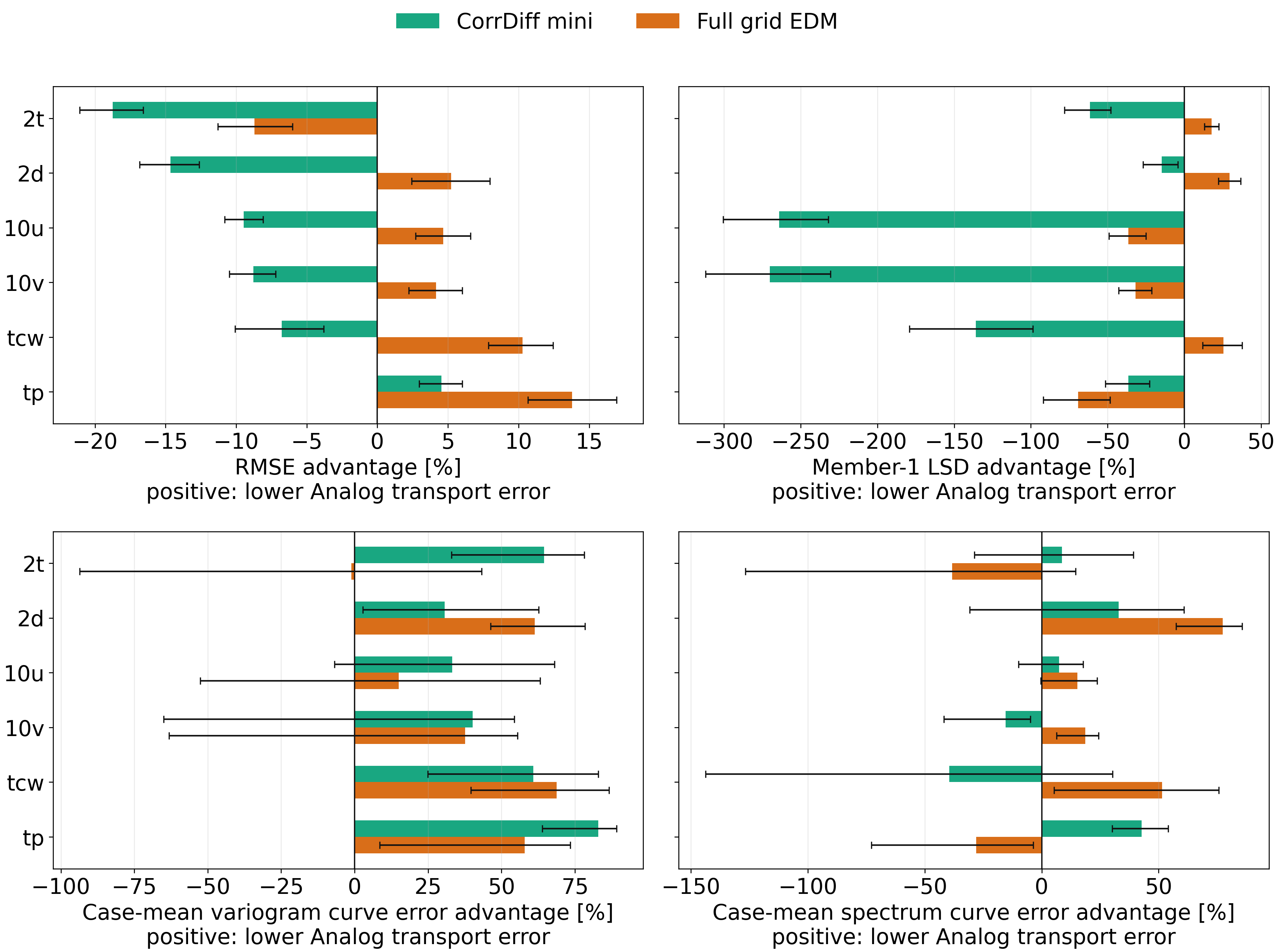}
\caption{Contextual intercomparison on 50 validation cases. Positive values favour analog transport. Bars show relative error reductions; whiskers show paired 95\% case-bootstrap intervals. Field labels are 2-m temperature (2t), dew point (2d), 10-m wind components (10u, 10v), total column water (tcw) and six-hour precipitation (tp).}
\label{fig:mini}
\end{figure}

Figure~\ref{fig:mini} compares six shared surface and column outputs on the validation sample. The local eight-variable CorrDiff-mini model,
trained on the same IFS/MEPS data, has lower ensemble-mean RMSE for the five
atmospheric outputs; analog transport has lower precipitation RMSE.
Against the available full-grid EDM member, the analog ensemble mean has
lower RMSE for all shared outputs except 2-m temperature. This comparison
uses the common valid footprint. The released-checkpoint comparison above
uses member 1 from each system.

Member-wise log-spectral distance (LSD) favours CorrDiff-mini for all six
outputs. Against full-grid EDM, analog transport has lower LSD for
temperature, dew point and total column water, and higher LSD for both wind
components and precipitation. Analog transport more closely reproduces the
case-mean variograms for dew point, total column water and precipitation
against both comparators, and for temperature against CorrDiff-mini.
The corresponding 95\% intervals exclude zero. Wind variogram differences
remain uncertain against both models.

The spectrum errors give a mixed ordering. Analog transport has
lower precipitation spectrum error than CorrDiff-mini, and lower errors
for dew point, meridional wind and total column water than full-grid EDM.
CorrDiff-mini better reproduces the meridional-wind spectrum, and full-grid
EDM better reproduces the precipitation spectrum. The remaining spectral
comparisons have intervals that include zero.
\FloatBarrier
\subsection{Low-cost 81-field ensemble generation}
The full downscaling model contains approximately 150 million parameters: about 27 million in the deterministic downscaler, 58 million in the atmospheric latent transport, 39 million in the precipitation latent transport and 23 million in the decoder. Archive construction and top-64 retrieval are nonparametric.
The training wall time totals approximately 12~h on one NVIDIA H200 GPU (Table~\ref{tab:stages}). Decoder time includes base reconstruction training and the humidity and precipitation specializations. The atmospheric transport is trained with the deterministic downscaler fixed, and the decoder is held fixed for transport evaluation.
\begin{table}[t]
\centering\tablefont
\begin{tabular}{lr}
\toprule
Selected training stages & H200 wall time\\
\midrule
Deterministic downscaler & 2 h 02 min\\
Atmospheric transport & 3 h 31 min\\
Grid decoder & 4 h 26 min\\
Precipitation transport & 2 h 15 min\\
\midrule
Selected-stage total & 12 h 14 min\\
\bottomrule
\end{tabular}
\caption{Training wall times for the selected model chain on one H200.}
\label{tab:stages}
\end{table}

For scale, the communicated training estimate for the CERRA/CARRA full grid EDM downscaler is approximately 60,000 GPU-hours. The published four-field Taiwan CorrDiff experiment used approximately 21,500 H100 GPU-hours \citep{mardani2025corrdiff}. The released REA6 checkpoint evaluated here has an unreported training budget. Hardware, output scope and training objectives differ across these systems (see Table~\ref{tab:training}).
\begin{table}[t]
\centering\tablefont
\begin{tabularx}{\linewidth}{@{}L{2.5cm}L{2.8cm}L{2.8cm}Y@{}}
\toprule
Model & Training & H200 inference & Native outputs\\
\midrule
Analog transport & 12 h 14 min (H200) & 12.3 s per member & 80 atmospheric fields + precipitation\\
Released CorrDiff & Unreported & about 5.8 s per sample & 45 fields\\
CorrDiff Taiwan & $\approx21{,}500$ H100 GPU-h & Unreported & 4 fields\\
Full grid EDM & $\approx60{,}000$ GPU-h & 30--60 s per timestamp & 16 surface fields\\
\bottomrule
\end{tabularx}
\caption{Training and inference budgets with each system's native output scope. Analog transport and released CorrDiff inference were measured on H200; full grid EDM figures were communicated to the author.}
\label{tab:training}
\end{table}

The 50-date, 20-member configuration benchmark produced 1,000 regional states in 3.4 H200 hours: 12.3~s per 81-field member, about 4~min for a 20-member valid time, and 13.5~GB peak allocation. This is the inference timing benchmark for the selected configuration. The released CorrDiff checkpoint averages about 5.8~s per sample on the same GPU and about 17~s per case including model loading, ERA5 access and regridding. Full grid EDM is reported at 30--60~s per valid time on an H200, depending on domain and configuration. The complete 20-member regional ensemble runs on one H200.

\subsection{Frozen-weight transfer beyond the training region}
\begin{figure}[t]
\centering
\includegraphics[width=\linewidth]{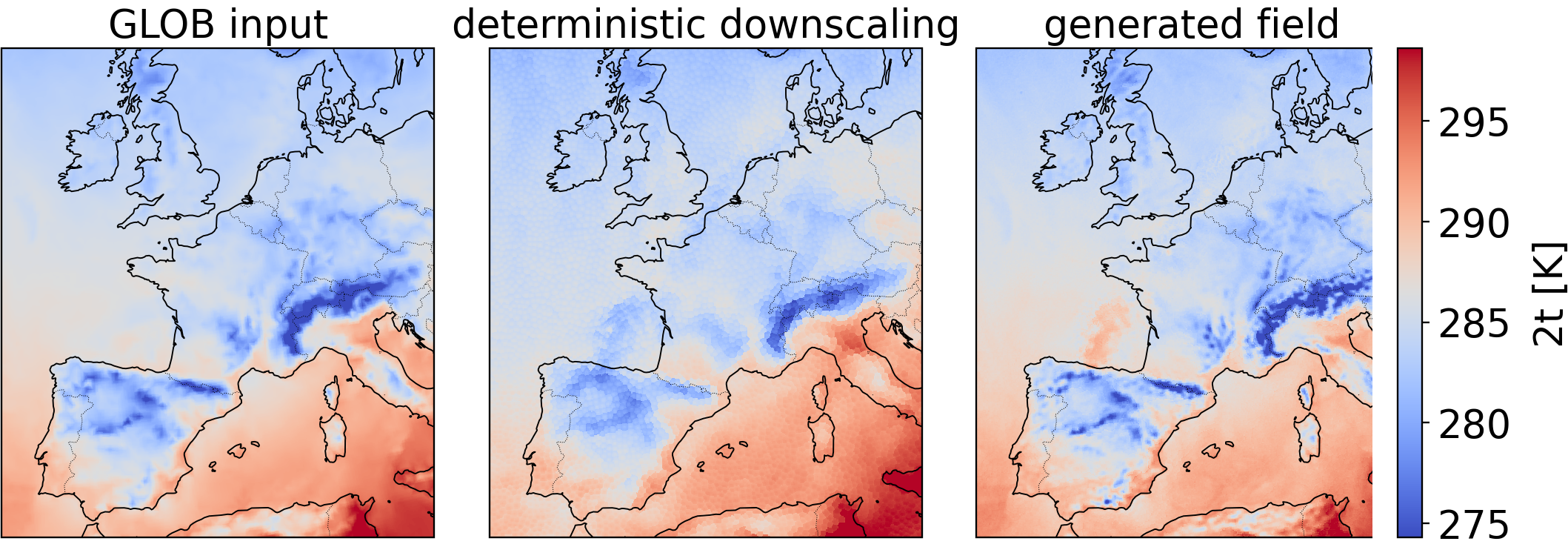}
\caption{France-centred 2-m temperature for the frozen-weight execution example, in kelvin. Panels show N320 IFS input, the deterministic downscaler and a generated field. The system uses Nordic-trained weights, France-specific static fields and an O1280-derived historical source bank. The common colour scale shows the deterministic contribution and the additional generated structure.}
\label{fig:france}
\end{figure}

The France experiment tests execution outside the Nordic training geography. After replacing coordinates, static fields and the analog bank, the unchanged transport and decoder produce 20 members of all 81 fields over $1{,}014{,}481$ points. Figure~\ref{fig:france} includes the frozen deterministic downscaler, making the increment supplied by the stochastic system visible.
The generated field follows the broad thermal contrast across France and the Alps and adds local temperature structure to the deterministic field. This case demonstrates execution of the connected latent representation on a translated grid with fixed transport and decoder weights. Quantitative skill over France remains unmeasured because an archive of aligned regional targets was unavailable for verification.

\section{Discussion}
The model proposed in this study produces regional ensembles of the atmospheric column, surface
weather and precipitation at a modest computational cost. The selected
staged training workflow requires approximately 12~h on one H200,
and each 81-field member takes about 12~s to generate. These outputs cover
12 pressure levels and more than one million regional grid points.
The PCA--mesh representation makes this breadth affordable by concentrating
the generative calculation in a latent space that is 12.5 times smaller
than the atmospheric output grid.

The 2023 validation comparisons show competitive quality on common fields
against full-grid models, including an EDM system trained with a
substantially larger computational budget. Analog transport performs
particularly well on precipitation RMSE and case-mean variograms. The resulting trade-off is useful for applications
requiring ensembles throughout the atmospheric column: the latent system
combines broad output coverage with good performance on several pointwise
and spatial diagnostics at a cost compatible with routine generation on
one GPU.

The individual members also retain visually plausible regional weather
structures. Temperature follows the mountain and coastal contrasts, and
wind and precipitation form continuous features across the domain.
Members contain local gradients and fine-scale variability that are
smoothed in the ensemble mean. Their spectra and variograms quantify
this recovery of spatial variability. The humidity transform ties its
physical bound to the generated temperature, and the precipitation
transform enforces nonnegative amounts. Analog initialization supplies
an existing multivariate residual pattern that the transport adjusts to
the current large-scale state and regional geography.

The latent construction also provides a basis for further regional
modelling. A shared factor representation, a connected mesh and a
latent-to-grid decoder allow models to operate on compressed residual
fields while retaining access to the native regional grid. The frozen-weight France example demonstrates that the same
transport and decoder can operate with new coordinates, static fields
and an IFS-derived historical bank, supporting the use of this
representation across regional domains.

\section{Conclusions}
We introduced an analog-initialized latent transport system for probabilistic
downscaling from N320 IFS to a 2.5-km regional grid. A deterministic downscaler
first supplies the predictable state. Historical residuals from similar IFS
situations then initialize 20 ensemble members, which are adjusted by a learned
probability flow in a 12.5-fold compressed spatial--factor latent and decoded
over the full domain.

The source distribution experiment measures how the initial distribution and its fitted transport affect the final output. Matched analog initialization improves point and spatial errors over Gaussian initialization and yields lower spectral error than covariance matching. The selected system improves CRPS over deterministic MAE for all 80 atmospheric channels and precipitation on test cases. Its selected training stages require approximately 12~h on one H200, with published or communicated budgets for the full-grid reference systems in the tens of thousands of GPU-hours.

The same frozen transport and decoder produce the 81-field product over France after replacing regional inputs. Analog initialization and whole-domain latent transport offer an efficient route from global analyses to regional ensembles, with explicit diagnostics of spatial variability, physical dependence and remaining calibration errors.

\acknowledgments
The author acknowledges the providers of IFS, MEPS and the static datasets,
and the developers of the released NVIDIA CorrDiff model and the Anemoi
framework. The author thanks Manuel Carrer from the Copernicus C3S2\_361b regional
downscaling project \citep{stoylen2026carra_cerra_ml} for insights and
information about the deployed model.
The author declares no conflicts of interest.
\datastatement
The released CorrDiff model is available from NVIDIA \citep{nvidia2026cosmo}; Earth2Studio and Anemoi provide public software interfaces. ERA5 is available through the Copernicus Climate Data Store \citep{hersbach2020era5}. The operational IFS and MEPS archives used here are held at the Norwegian Meteorological Institute and remain subject to their providers' access and redistribution conditions. 

\clearpage
\begin{appendix}[A]
\appendixtitle{Static context and France-native source adaptation}
The Nordic physiography map is constructed in order: sea where land fraction
is below 0.5; dense city where urban fraction is at least 0.10; complex
terrain where elevation is at least 300 m, roughness at least 50 m, slope
at least 10 m km$^{-1}$ or the absolute MEPS-minus-IFS elevation difference
at least 50 m; dense forest where forest fraction is at least 0.80 after
the previous exclusions; and ordinary land elsewhere. Continuous terrain
and land--sea context are retained alongside these classes.
To run the model over France, the grid shape, nominal spacing, 100,000-node mesh connectivity
and trained weights are kept frozen, while coordinates, static fields and the historical analog bank change.
France-specific terrain comes from the European digital terrain product
identified in the provenance package, resampled to approximately 510 m.
The IFS land--sea mask and ECOCLIMAP-SG 300-m land cover
\citep{cnrm2026ecoclimap} replace region-specific static files. Urban cover
classes 24--33 and forest classes 7--14 and 22 are averaged to regional
fractions. Inference over France applies the city threshold above and
classifies complex terrain using elevation or slope, followed by forest and
ordinary land. Static coverage is complete on the translated grid.

The current-time input over France is N320 IFS. The frozen deterministic
downscaler produces the regional prediction from which we construct the
retrieval descriptor $e(x)$. We construct
the historical analog bank from paired N320 and O1280 IFS states. For each
historical date $j$, the frozen deterministic downscaler predicts the
regional state from N320 IFS and French static fields. The difference
between the corresponding O1280 fields and this prediction defines the
source residual, which we standardize and encode as
\begin{equation}
F_{0,j}^{D'}=E_{D'}\left[
\left\{T(x_{O1280,j})-T(D_\theta(x_{N320,j},s_{D'}))\right\}
A^{-1}B\right],
\label{eq:francesource}
\end{equation}
where the O1280 fields and deterministic prediction are represented on
the same French regional grid, and $A=\operatorname{diag}(a_v)$. The resulting bank is
\begin{equation}
\mathcal A_{D'}=
\left\{\left(e(x_{N320,j}),F_{0,j}^{D'}\right)\right\}_{j=1}^{128}.
\label{eq:francebank}
\end{equation}
The N320 member of each historical pair supplies the retrieval context, while
the O1280 member supplies the higher-resolution side of the stored residual.
Top-64 retrieval produces 20 distinct source fields. We use the fixed
per-factor moment calibration from Nordic training-factor statistics. At evaluation
time, N320 IFS supplies the model input; the O1280 states contribute to the precomputed historical bank.

Transport weights, deterministic model, channel basis and decoder remain
the ones fitted on the Nordic domain; regional adaptation is carried by the
O1280-derived historical source fields, their paired N320 retrieval contexts,
the current N320 IFS input and high-resolution static fields.
Figure~\ref{fig:france} in the main text illustrates the resulting adaptation
to French geography.
\end{appendix}
\FloatBarrier
\Needspace{6\baselineskip}
\begin{appendix}[B]
\appendixtitle{Humidity and precipitation coordinates}
Specific humidity is represented relative to saturation so that its physical
bound follows the generated temperature. Let $\vartheta_p$ and
$\widehat\vartheta_p$ denote target and generated temperature at pressure
level $p$. The transformed and reconstructed humidity are
\begin{align}
h_p&=\operatorname{logit}\left[
\operatorname{clip}\left\{
\frac{q_p}{c_p q_{\rm sat}(\vartheta_p,p)},
\epsilon,1-\epsilon
\right\}\right],\\
\widehat q_p&=c_p q_{\rm sat}(\widehat\vartheta_p,p)
\operatorname{sigmoid}(\widehat h_p).
\label{eq:humidity}
\end{align}
We set $\epsilon=10^{-5}$. The saturation function uses a mixed-phase Buck water--ice formulation
\citep{buck1981}; caps $c_p$ and residual scales are fitted on training data.
The inverse uses the generated temperature at the same pressure level.
Six-hour accumulated precipitation is represented on the 100,000-node mesh as
\begin{equation}
z=\log\left\{1+\frac{\max(\mathrm{tp},0)}{0.1\,\mathrm{mm}}\right\}.
\end{equation}
The inverse is
\begin{equation}
\widehat{\mathrm{tp}}=\min\!\left\{268.67~\mathrm{mm},\;
0.1~\mathrm{mm}\,[\exp\{\max(\widehat z,0)\}-1]\right\}.
\end{equation}
Nonpositive generated coordinates produce exact zero. The upper bound is the frozen training maximum, used as a numerical support guard. It clips about 3 in every 100,000 generated member-grid values on the 96-case test. All ensemble-mean values lie below this bound. The precipitation contribution to the joint decoder loss combines pointwise, spatial-gradient and spectral terms.
\end{appendix}
\FloatBarrier

\Needspace{6\baselineskip}
\begin{appendix}[C]
\appendixtitle{Network training and source distribution experiments}
\label{app:training}
The deterministic downscaler, latent transports and joint grid decoder are trained
in stages. The deterministic downscaler uses width 256, separate IFS and
regional forward mappers, a two-layer hidden-mesh processor with four
attention heads, and a regional backward mapper with a latent skip. Its graph
uses 12-neighbour IFS-to-mesh, 3-neighbour regional-to-mesh, 8-neighbour
mesh-to-mesh and 1-neighbour mesh-to-grid connections. Regional inputs contain
11 target-independent elevation, land--sea, coordinate, calendar, local-time
and insolation channels. We freeze its final 10,000-update checkpoint for
residual construction and all subsequent training.

The atmospheric velocity model operates on 65 factors at 100,000 nodes. Its
processor has eight graph-transformer layers, width 512, eight attention heads
and two computational chunks. Table~\ref{tab:optimization} gives the selected
training schedules. All three listed networks use batch size one and bf16
mixed precision. The deterministic objective is weighted, area-normalized
mean-square error. Its AdamW momentum coefficients are $(0.9,0.95)$, with
unit-norm gradient clipping. Atmospheric transport uses weighted velocity
mean-square error and AdamW with the same momentum coefficients, weight decay
0.1, numerical constant $10^{-7}$ and unit-norm gradient clipping. Precipitation
transport adds graph-gradient and spectral losses with weights 0.20 and 1.0.

\begin{table}[t]
\centering\tablefont
\begin{tabular}{lrrr}
\toprule
Selected network & Updates & Peak learning rate & Warm-up updates\\
\midrule
Deterministic downscaler & 10,000 & $6.25\times10^{-5}$ & 1,000\\
Atmospheric transport & 3,000 & $1.5\times10^{-5}$ & 100\\
Precipitation transport & 10,000 & $5\times10^{-5}$ & 500\\
\bottomrule
\end{tabular}
\caption{Optimization schedules for the selected system. Learning rates
follow cosine decay to $3\times10^{-7}$ after warm-up. The source distribution controls
use the separate 10,000-update protocol described in Section~\ref{sec:source-controls}.}
\label{tab:optimization}
\end{table}

The grid decoder lifts the factors and 12 mesh/static features
to width 256 and applies a graph-transformer backward mapper. Its local
convolutional component has 96 channels and four depthwise residual blocks,
with dilations 1, 2, 4 and 8. The humidity specialization is a
189-to-256-to-12 multilayer perceptron with the saturation-logit inverse in
Appendix~B. The decoder is trained on encoded target
residuals and then frozen for transport evaluation.
\end{appendix}
\FloatBarrier
\bibliographystyle{ametsocV6}
\bibliography{bibli}
\end{document}